\documentclass[twocolumn,journal,10pt]{IEEEtran}
\usepackage{epsfig,amsfonts,psfrag}
\usepackage[nolist]{acronym}
\usepackage{graphicx,cite,amssymb,amsmath,color}
\usepackage{psfrag}
\usepackage[dvipsnames]{xcolor}
\usepackage{mathtools}
\mathtoolsset{showonlyrefs}

\begin{document}
\begin{acronym}
\acro{CRDSA}{contention resolution diversity slotted ALOHA}
\acro{CSA}{coded slotted ALOHA}
\acro{IRSA}{irregular repetition slotted ALOHA}
\acro{IoT}{internet of things}
\acro{LEO}{low Earth orbit}
\acro{MAC}{medium access control}
\acro{NB-IoT}{narrowband-IoT}
\acro{NOMA}{non-orthogonal multiple access}
\acro{RA}{random access}
\acro{SA}{slotted ALOHA}
\acro{SIC}{successive interference cancellation}
\acro{TDMA}{time division multiple access}
\end{acronym}

\title{\huge Modern Random Access for Beyond-5G Systems: \\a Multiple-Relay ALOHA Perspective}
\author{Andrea Munari, Federico Clazzer%
\vspace{-1.5em}
\thanks{
Authors are with Institute of Communications and Navigation of the German Aerospace Center (DLR), 82234 Wessling, Germany (e-mail: \{Andrea.Munari, Federico.Clazzer\}@dlr.de).} 
\thanks{This work has been accepted for publication at the 3rd Balkan Conference on Communications and Networking, 2019.}}
\maketitle
\thispagestyle{empty} \setcounter{page}{0}


\begin{abstract}
Modern random access protocols are emerging as an efficient yet simple solution for arising \ac{IoT} applications in upcoming beyond-5G systems. In this context, both terrestrial and non-terrestrial scenarios can benefit from the presence of multiple low-complexity receivers that act as relays, collecting packets from users and forwarding them towards a central unit. To gain insights on the potential of these setups, we investigate a two-tier \ac{SA} multiple-relay system under an erasure channel model. We derive exact end-to-end throughput expressions for an arbitrary number of receivers, and complement our study by discussing the impact of channel impairments. The non-trivial outcome that adding relays is not always beneficial is highlighted and thoughtfully discussed.
%
\end{abstract}

{\pagestyle{empty}

\newcommand{\nRx}{\ensuremath{\mathsf K}}
\newcommand{\tru}{\ensuremath{\mathsf S}}
\newcommand{\truB}{\ensuremath{\tilde{\tru}}}
\newcommand{\maxTru}{\ensuremath{{\tru^*}}}
\newcommand{\truSA}{\ensuremath{\mathsf S_{\mathsf{sa}}}}
\newcommand{\truU}{\ensuremath{\mathsf S_{\mathsf u}}}
\newcommand{\truD}{\ensuremath{\mathsf S_{\mathsf d}}}
\newcommand{\load}{\ensuremath{\mathsf G}}
\newcommand{\peras}{\ensuremath{\varepsilon}}
\newcommand{\perasU}{\ensuremath{\peras_{\mathsf u}}}
\newcommand{\perasD}{\ensuremath{\peras_{\mathsf d}}}
\newcommand{\pforw}{\ensuremath{\delta}}
\newcommand{\NTx}{\ensuremath{\mathsf N}}
\newcommand{\nTx}{\ensuremath{n}}
\newcommand{\psU}{\ensuremath{\mathsf p_{\nTx}}}
\newcommand{\psD}{\ensuremath{\mathsf q_{\nTx}}}
\newcommand{\ancF}{\ensuremath{\mathcal H}} 
\section{Introduction}
\label{sec:intro}

\IEEEPARstart{T}{he} day-by-day unfolding of new \ac{IoT} applications poses challenges that are only partially addressed by current 5G solutions, and shall be of primary concern in beyond-5G communication systems. Attaining high energy- and spectral-efficiency communications, with a vast user population that sporadically generates small amount of data following, at times, unpredictable activation patterns is a very demanding task. In this context, approaches implemented in current standards and based on orthogonal allocation of resources become inefficient, due to the high cost of overhead undergone as the transmitter population grows and when high flexibility is required. An appealing alternative is to employ \ac{RA} protocols. In their simplest form, these schemes foresee nodes to access the shared medium without coordination, thus sparing the need for costly resource-grant procedures.

Nowadays, medium access solutions addressing the \ac{IoT} ecosystem both in the licensed, e.g. \ac{NB-IoT} and its evolution \cite{TS38.321}, and in the unlicensed spectrum, e.g. SigFox~\cite{SigFox}, LoRaWAN~\cite{LoRa}, still largely rely on the classical ALOHA paradigm of the early 1970s~\cite{Abramson1970,Roberts72:ALOHA}. In the last decade, the adoption of multi-user detection in \ac{RA} protocols has generated a wave of research activity collectively labeled as \emph{modern random access}~\cite{Berioli2016}, drastically improving their performance. In the original configuration, such protocols rely on the use of time diversity \--- having users transmit multiple copies of their packets \--- and \ac{SIC} at the receiver~\cite{Casini2007}. Extensions adopting an optimised number of repetitions~\cite{Paolini15:MAG} have been shown to asymptotically achieve the performance of scheduled access under ideal channel conditions~\cite{Narayan12:BoundIRSA}.

Notwithstanding the remarkable performance enhancements, these approaches entail modifications at the transmitter side compared to ALOHA that may hinder a straightforward application to (beyond-)5G and that may be unaffordable for low-cost and low-complexity devices. In light of this, we take a different lead and investigate the potential of spatial diversity, studying a system in which the transmitters access the medium following the \ac{SA} protocol while reception is attempted at different positions. We do not consider here the case in which a single receiver is equipped with multiple antennas~\cite{Zorzi98:ALOHA,LaMaire96:Diversity}. Instead, as originally pioneered in~\cite{Corson1993}, we leverage on having a set of disjoint receivers equipped with single antenna, attempting the collection of sent data units. Having in mind \ac{IoT} applications, where information often has to reach a single collection point for further processing, we complement the topology assuming that receivers (relays) forward decoded packets to a common sink. This two-tier setup has been investigated in~\cite{Munari19_arXiv} assuming orthogonal resource allocation for the receivers-to-sink links, deriving insightful bounds and practical forwarding policies.\footnote{Multiple-relay in conjunction with time-diversity was studied in \cite{Jakovetic15:TCOM} assuming coordination among receivers.} In this work, instead, the relays contend for the medium following a \ac{SA} policy as well. Such a model fits well, among others, two scenarios relevant for beyond-5G systems: ultra-dense heterogenous networks and \ac{LEO} mega-constellations. In heterogenous networks, a macro base-station (sink) is complemented by the presence of many smaller base-stations (possibly with limited processing power~\cite{An17:RAHetNet, Pan18:CloudCRAN}) that relay information from the users to increase capacity and reduce latency~\cite{Wang14:KeyTec5G, Condoluci15:HetNet}. On the other hand, non-terrestrial communications represent an interesting use case, especially in their embodiment as mega-constellations, e.g. OneWeb \cite{OneWeb}, Amazon Kuiper Project, and SpaceX Starlink. In these systems, worldwide coverage is provided by the presence of hundreds, or more, small \ac{LEO} satellites. The probability of being in visibility of multiple satellites from any location on Earth becomes very high, calling for the exploitation of spatial diversity for data reception. For ground terminals, indeed, satellites can be seen as multiple receivers that collect data later to forward it to a gateway.

Inspired by these applications, we study in this paper the performance of a two-tier \ac{SA} multiple-relay system. Resorting on the simple erasure channel model, we derive the exact close form expression of the end-to-end throughput for an arbitrary number of relays. Allowing receivers to probabilistically forward to the sink what collected, we discuss the key trends that emerge, highlighting the role of channel impairments. Non-trivial tradeoffs emerge, revealing how increasing the number of receivers may be not always beneficial, as the optimal cardinality of the relay set is driven by channel conditions and load.

\section{System Model}
\label{sec:sysModel}

Throughout our discussion we focus on the topology of Fig.~\ref{fig:topology}, where an infinite population of users generate traffic in the form of data packets addressed to a common sink. Time is divided in slots (or time units) of equal duration, with all devices being slot-synchronous. No direct link between users and sink is available, and communications take place in two steps.
During the former, users access the shared wireless \emph{uplink} channel obeying a \ac{SA} protocol, and attempt data delivery towards a set of $\nRx$ receivers (or relays). Following a well-established model, the number of users transmitting a data unit over a slot is described by a Poisson random variable $\NTx$, whose intensity $\load$ [pk/slot] is referred to as channel load. The links between terminals and receivers are characterised as on-off fading channels \cite{OnOff2003}, so that a transmitted packet either reaches a relay with probability $1-\perasU$, or is erased (i.e.  brings no power contribution) with probability $\perasU$. For the sake of mathematical tractability, erasure events are assumed to be i.i.d. over time as well as for each transmitter-receiver pair within a slot. No multi-user detection capabilities are available, and collisions are regarded as destructive, while singleton packets are correctly decoded.  Accordingly, a relay successfully retrieves information when only one of the $\NTx = \nTx$ transmitted data units arrives unfaded, i.e. with probability $\psU := \nTx \,(1-\perasU) \, \perasU^{\nTx-1}$. Removing the condition on $\NTx$, the successful decoding probability at a receiver readily follows:
\begin{align}
\truSA = \sum_{\nTx=0}^{\infty}  \frac{\load^\nTx \, e^{-\load}}{\nTx!} \cdot  \psU = \load (1-\perasU) \, e^{-\load (1- \perasU)}
\label{eq:truSA}
\end{align}
which corresponds to the throughput of a \ac{SA} link with erasures.

The setup is then complemented by an orthogonal \emph{downlink} channel, shared by relays through a \ac{SA} policy to forward collected information towards the sink. Specifically, each receiver operates in out-of-band full-duplex mode, and, upon decoding a data unit from a user, it independently decides whether to transmit it on the downlink in the subsequent slot (with probability \pforw) or to discard it (with probability $1-\pforw$).\footnote{In other words, no buffering is performed at relays, so that a received packet is either immediately forwarded to the sink or never so. For a study on systems where buffering is allowed, the interested reader is referred to \cite{Munari19_arXiv}.} Relay-to-sink connections are modelled as i.i.d. on-off fading links with erasure probability $\perasD$, and a slot leads to data retrieval only if a single packet reaches the sink.

We characterise system performance by means of the \emph{end-to-end throughput} $\tru$, defined as the average number of packets received at the sink per slot. The metric clearly depends on the number of available receivers \nRx\ and, in spite of its simplicity, captures the fundamental tradeoff of the two-tier topology under study. Indeed, while deploying a larger relay population improves the probability of collecting packets sent by the users, it also raises the contention over the downlink channel, creating a potential bottleneck for successful forwarding to the sink. An upper bound to the achievable performance of the system as well as a closed-form expression of the end-to-end throughput will be presented in the following section.
\begin{figure}
\centering
\includegraphics[width=.8\columnwidth]{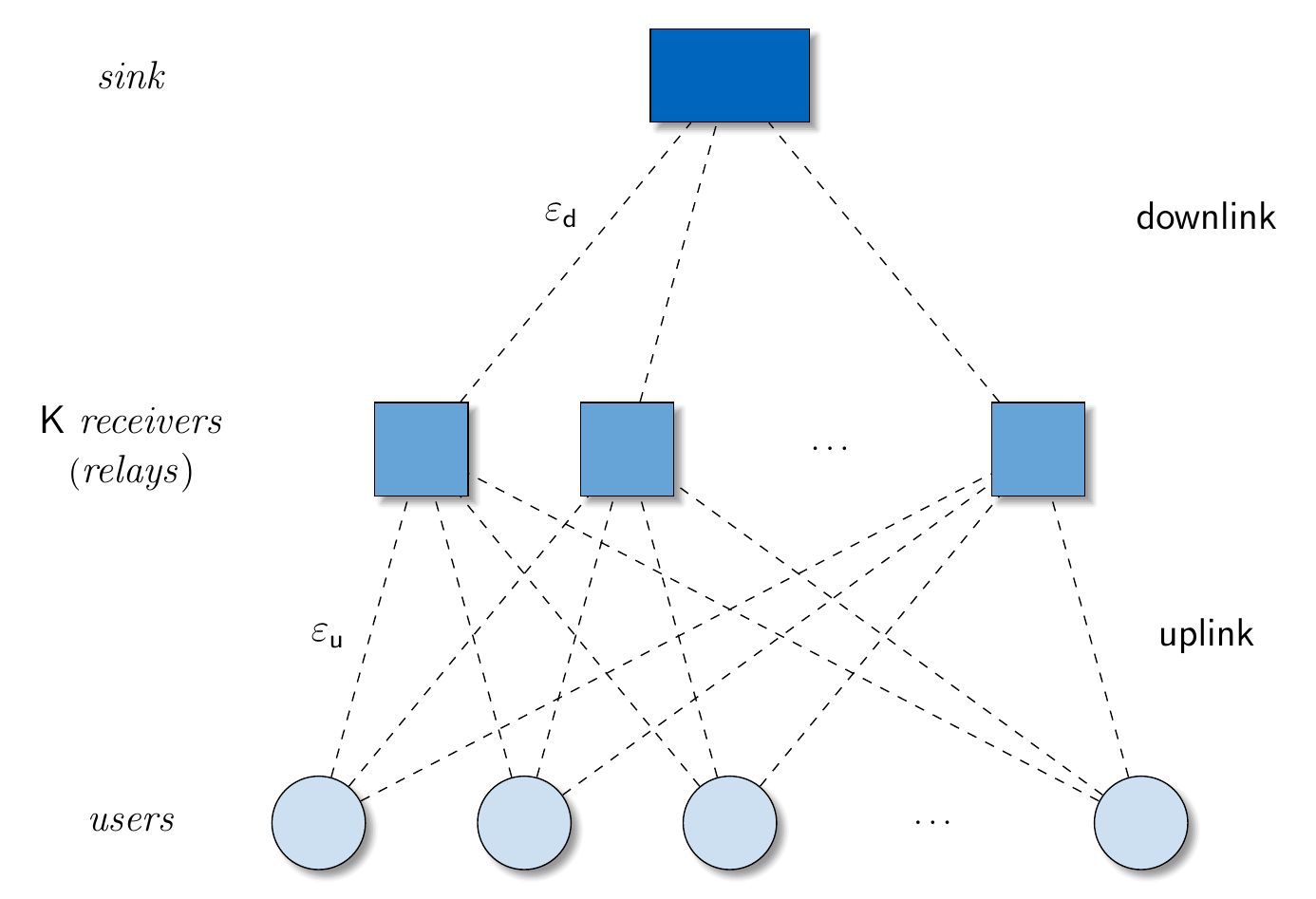}
\caption{Reference system topology.}
\label{fig:topology}
\end{figure} 

\section{Analysis and Results}
\label{sec:analysis}

To derive the end-to-end throughput of the system, consider first the event of having a packet sent on the uplink reach the sink via one of the relays. Assuming $\NTx=\nTx$ users transmitted, this occurs when the receiver successfully decodes one of the users' data units, forwards it, and the packet is not erased over the downlink, with overall probability $\psD := \psU \pforw \,(1-\perasD)$.
By virtue of the independence of erasures events, the number of incoming downlink data units over a slot follows a binomial distribution of parameters $(\nRx,\psD)$ and, recalling that collisions are regarded as destructive, information is retrieved only when a single packet reaches the sink, i.e. with probability \mbox{$\nRx \psD (1-\psD)^{\nRx-1}$}. The end-to-end throughput can then be formulated as
\begin{equation}
\tru = \sum_{\nTx=0}^{\infty} \frac{\load^\nTx e^{-\load}}{\nTx!} \cdot \nRx \, \psD (1-\psD)^{\nRx-1}
\label{eq:truSum}
\end{equation}
leading to the closed for expression in \eqref{eq:tru} at the top of next page, whose derivation is reported in App.~\ref{app:tru}, and where the ancillary function $\ancF_m(x)$ is defined recursively as
\begin{figure*}[t]
\normalsize
\begin{equation}
\tru = \sum_{\ell=0}^{\nRx -1} (-1)^\ell \, \nRx \, {\nRx-1 \choose \ell} \left[\frac{\pforw \, (1-\perasU) (1-\perasD)}{\perasU}\right]^{\ell+1}  e^{-\load} \cdot \ancF_{\ell+1}\!\left(\load \,\perasU^{\ell+1}\right)
 \label{eq:tru}
\end{equation}
\hrulefill
\end{figure*}
\begin{align}
\begin{split}
\ancF_0(x) &= e^x\\
\ancF_m(x) & = \sum_{\ell=0}^{m-1} {m-1 \choose \ell} \ancF_\ell (x) \quad m\geq1 .
\end{split}
\label{eq:ancFunc}
\end{align}
The result in \eqref{eq:tru} offers a compact characterisation of the system, conveniently capturing for any number of relays the role played by all key parameters. In order to better gauge the impact of having a random access dowlink channel, we complement our analysis deriving a relevant benchmark for the achievable performance. Specifically, let us consider an ideal system in which the sink retrieves a single data unit as soon as at least one of the relays decodes a packet over the uplink. Given a \ac{SA} operated uplink, this configuration reaps the best possible performance when no multi-user detection capabilities are available at the sink and receivers are not allowed to buffer incoming data units, as it removes all sources of packet losses over the downlink (i.e. collisions and erasures). Indicating by $\truB$ the end-to-end throughput of such a system, so that $\tru < \truB < 1$, we prove in App.~\ref{app:truBound} that
\begin{equation}
\truB = 1- \sum_{\ell=0}^{\nRx} (-1)^\ell \, {\nRx \choose \ell} \left(\frac{1-\perasU} {\perasU}\right)^{\ell}  e^{-\load} \cdot \ancF_{\ell}\!\left(\load \,\perasU^{\ell}\right).
 \label{eq:truBound}
\end{equation}

\subsection{The $\nRx=2$ relay case}

\begin{figure}
\centering
\includegraphics[width=.9\columnwidth]{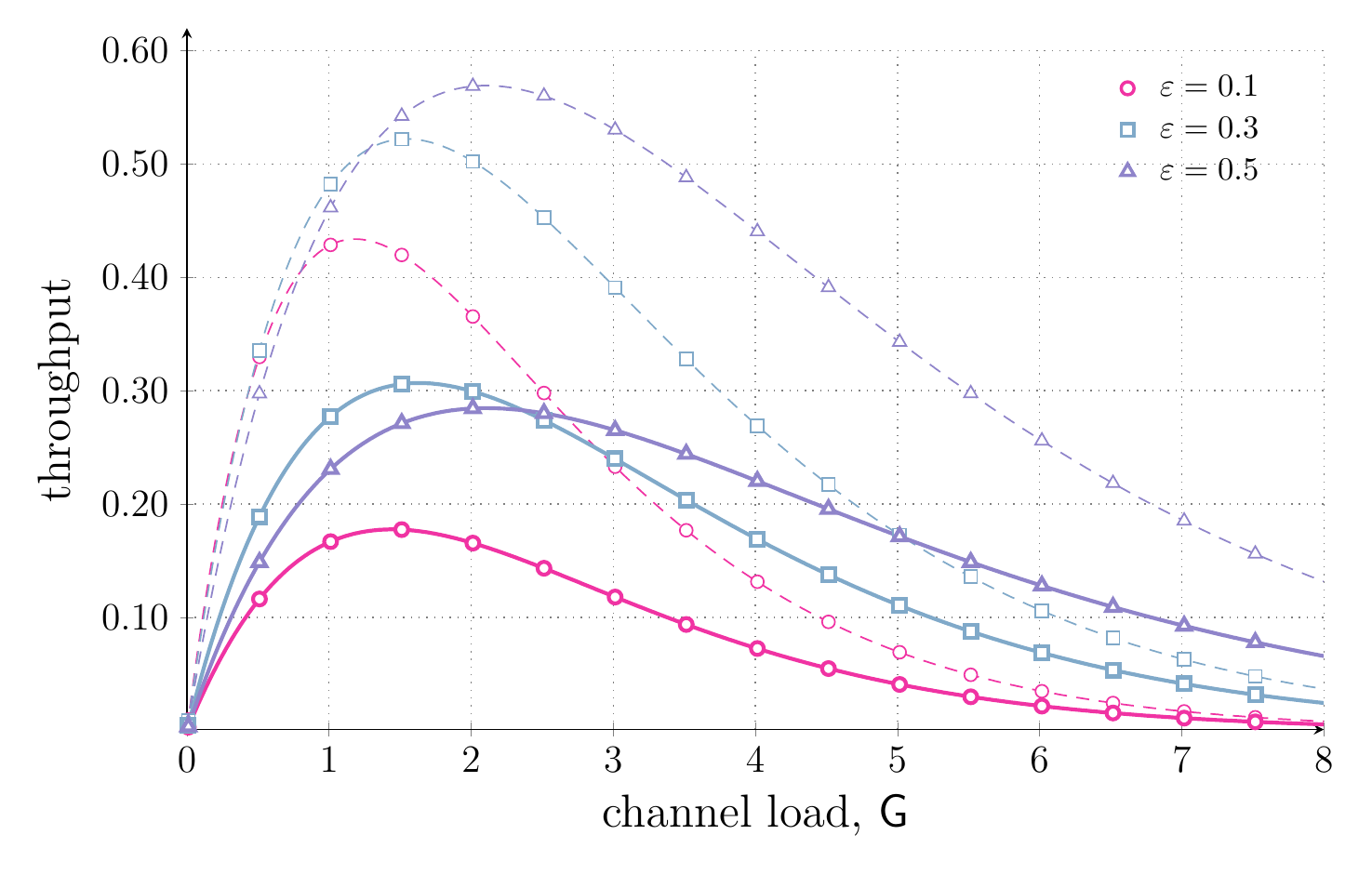}
\caption{End-to-end throughput vs channel load for different erasure rates $\perasU=\perasD:=\peras$. Solid lines report the trend of the system under study, whereas dashed lines the performance upper bound \truB.  $\nRx=2$, $\pforw=1$.}
\label{fig:tru2RxSymm}
\end{figure}

We start our discussion considering a simple yet practically relevant setup in which only two receivers are available. Assuming up- and downlink to be characterised by the same erasure rate \mbox{$\perasU = \perasD := \peras$}, Fig.~\ref{fig:tru2RxSymm} reports by solid lines the end-to-end throughput \tru\ when relays always forward what received (i.e. $\pforw=1$), and, by dashed lines, the performance upper bound \truB.\footnote{We recall that, by definition, the performance upper bound is not affected by erasures on the downlink.} Both metrics are depicted against the channel load \load.
Let us first focus on the bound, and recall that it is defined as the probability that at least one relay decodes a packet over a slot. In lightly loaded conditions ($\load \ll 1$), uplink channel impairments are the main driver, with better performance achieved for lower values of \peras. Conversely, when more users access the shared medium, erasures become beneficial in reducing the probability of collisions at each receiver, so that larger \truB\ are attained by increasing \peras. Such a trend significantly changes when the downlink is operated following a \ac{SA} policy and relay-sink connections are subject to erasures as well. Indeed, a higher probability for receivers to decode users' packets triggers a harsher contention in forwarding what collected, besetting the end-to-end throughput due to collisions at the sink. The effect clearly emerges from the plot at low loads, where the worst behaviour is experienced for $\peras=0.1$, and reveals an interesting tradeoff for intermediate values of \load, where the highest peak throughput is achieved for $\peras=0.3$ \--- corresponding to poorer uplink performance compared to the $\peras=0.5$ case.

\begin{figure}
\centering
\includegraphics[width=.97\columnwidth]{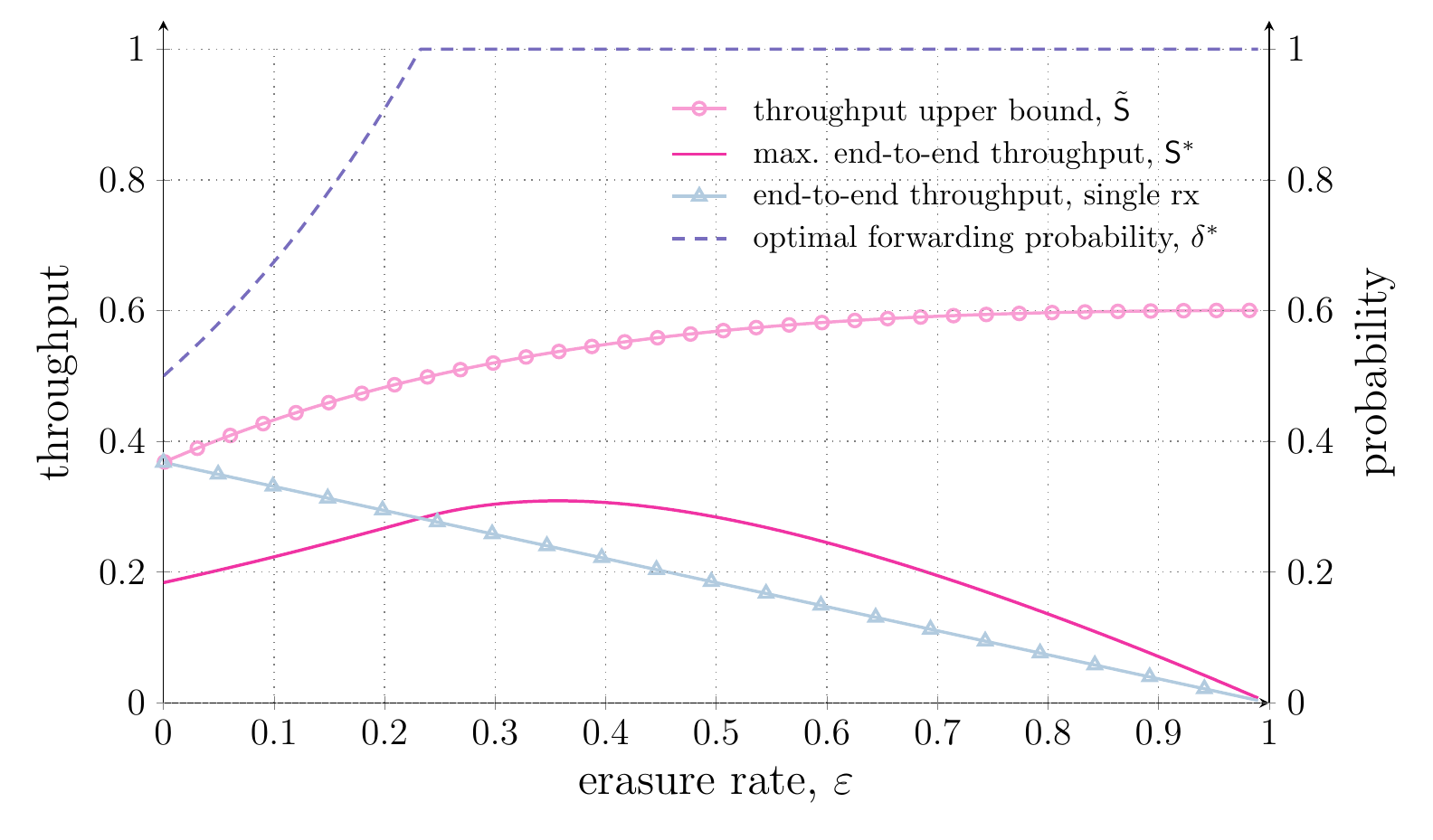}
\caption{Maximum achievable end-to-end throughput (solid line), and corresponding forwarding probability $\pforw^*$ (dashed line) vs channel erasure rate for $\nRx=2$. The circle-marked line reports the throughput upper bound \truB, whereas the triangle-marked one the end-to-end throughput in the presence of a single receiver. In all cases, $\perasU=\perasD:=\peras$ and $\load=1/(1-\peras)$.}
\label{fig:PeakTru2RxSymm}
\end{figure}
Fig.~\ref{fig:tru2RxSymm} highlights that the \ac{RA} downlink represents the system bottleneck, triggering the natural question on how to properly tune the forwarding probability so as to maximise the end-to-end performance and strike the proper balance between attempting delivery of collected information and limiting detrimental collisions at the sink. To delve into this issue, and to better isolate the relationship between \pforw\ and erasures, we fix the channel load, operating the system at $\load = 1/(1-\perasU)$,\footnote{As readily verified through \eqref{eq:truSA}, this choice maximises the throughput experienced by each relay, and is thus of practical relevance.} In such conditions, \eqref{eq:tru} takes the form
\begin{equation}
\tru = \frac{2 \pforw (1-\perasD)}{e} \cdot \left[ 1 - \pforw(1-\perasD) (1 -\perasU + \perasU^2)\, e^{-\perasU} \right]
\end{equation}
leading to a concave quadratic function of \pforw. The forwarding probability $\pforw^*$ that maximises the end-to-end throughput for an $(\perasU,\perasD)$ configuration can thus be computed by setting $\partial \tru /\partial \pforw = 0$ and recalling that $\pforw \in [0,1]$, to obtain
\begin{align}
\pforw^* = \min\left\{ 1, \,\frac{e^{\perasU}}{2(1-\perasD)(1-\perasU+\perasU^2)}\,\right\}.
\end{align}
The optimal throughput \maxTru\ that can be achieved using $\pforw^*$ readily evaluates to
\begin{equation}
\maxTru =
\begin{dcases}
\frac{e^{-1+\perasU}}{2(1-\perasU+\perasU^2)}                                                           & \pforw^* < 1 \\[.2em]
\frac{2 (1-\perasD)}{e} \cdot \left[ 1 - (1-\perasD) (1 -\perasU + \perasU^2)\, e^{-\perasU} \right]    & \pforw^* = 1
\end{dcases}
\label{eq:optTru}
\end{equation}
and is shown by a solid line in Fig.~\ref{fig:PeakTru2RxSymm} against the erasure rate for the special case $\perasU=\perasD=\peras$. For completeness, the plot also reports the corresponding values of $\pforw^*$ (dashed line), the throughput upper bound (circle-marked line), and the end-to-end throughput when a single relay is available (triangle-marked line). As discussed, for the load under consideration ($\load > 1$), uplink performance improves for larger values of \peras, leading to a monotonically increasing trend of \truB\ (which, by definition, is independent of \pforw). Indeed, for $\peras=0$ both relays always observe the same uplink outcome (either decoding if a lone packet was sent or not retrieving information), so that no benefit can be reaped out of receiver diversity. In turn, higher erasure rates favour a decorrelation of what relays observe over a slot, increasing the probability for at least one of them to successfully collect a data unit.\footnote{For \peras\ approaching $1$, the uplink behaviour can be approximated by considering relays observe two independent channels, so that, recalling \eqref{eq:truSA}, $\truB = (1-(1-\truSA)^2) = 1-(1-1/e)^2 \simeq 0.6$, as confirmed by Fig.~\ref{fig:PeakTru2RxSymm}.} Therefore, the optimal forwarding policy in the absence of erasures foresees each receiver drop the packet with probability $1-\pforw^*=1/2$, relying on its peer to deliver what collected to the sink. Instead, as $\peras$ increases it becomes more convenient for both relays to forward. The trend is apparent in Fig.~\ref{fig:PeakTru2RxSymm}, and continues up to the point where $\pforw^*$ saturates to $1$. In terms of optimal end-to-end throughput, this reflects into an increase in performance for up to moderate values of \peras, whereas, for strongly impaired channels, erasures on the downlink become dominant, and lead to a sharp decrease in \maxTru. More interestingly, Fig.~\ref{fig:PeakTru2RxSymm} enables a comparison between the $K=2$ configuration and the simplest setup in which a single relay is available. In the latter case, the receiver always forwards what decoded towards the sink, leading to an end-to-end throughput of \mbox{$(1-\peras) \,\truSA = (1-\peras) \, e^{-1}$} packets per slot. From the plot we observe that \--- for an ALOHA-operated downlink \--- the presence of an additional receiver turns out to be beneficial only for large enough erasure rates, as an outcome of both a diversity gain in the uplink and a stronger resiliency to channel impairments in the downlink. Conversely, when the relays experience highly correlated reception patterns (low \peras), the uncoordinated nature of their connections to the sink causes a bottleneck that outweighs the increased probability of retrieving packets from users. This result offers non-trivial insights on the two-layer system under study, and pinpoints the need to devise more advanced forwarding policies that reap the potential of multiple receivers when the downlink has to be operated following \ac{RA} procedures.
\begin{figure}
\centering
\includegraphics[width=.9\columnwidth]{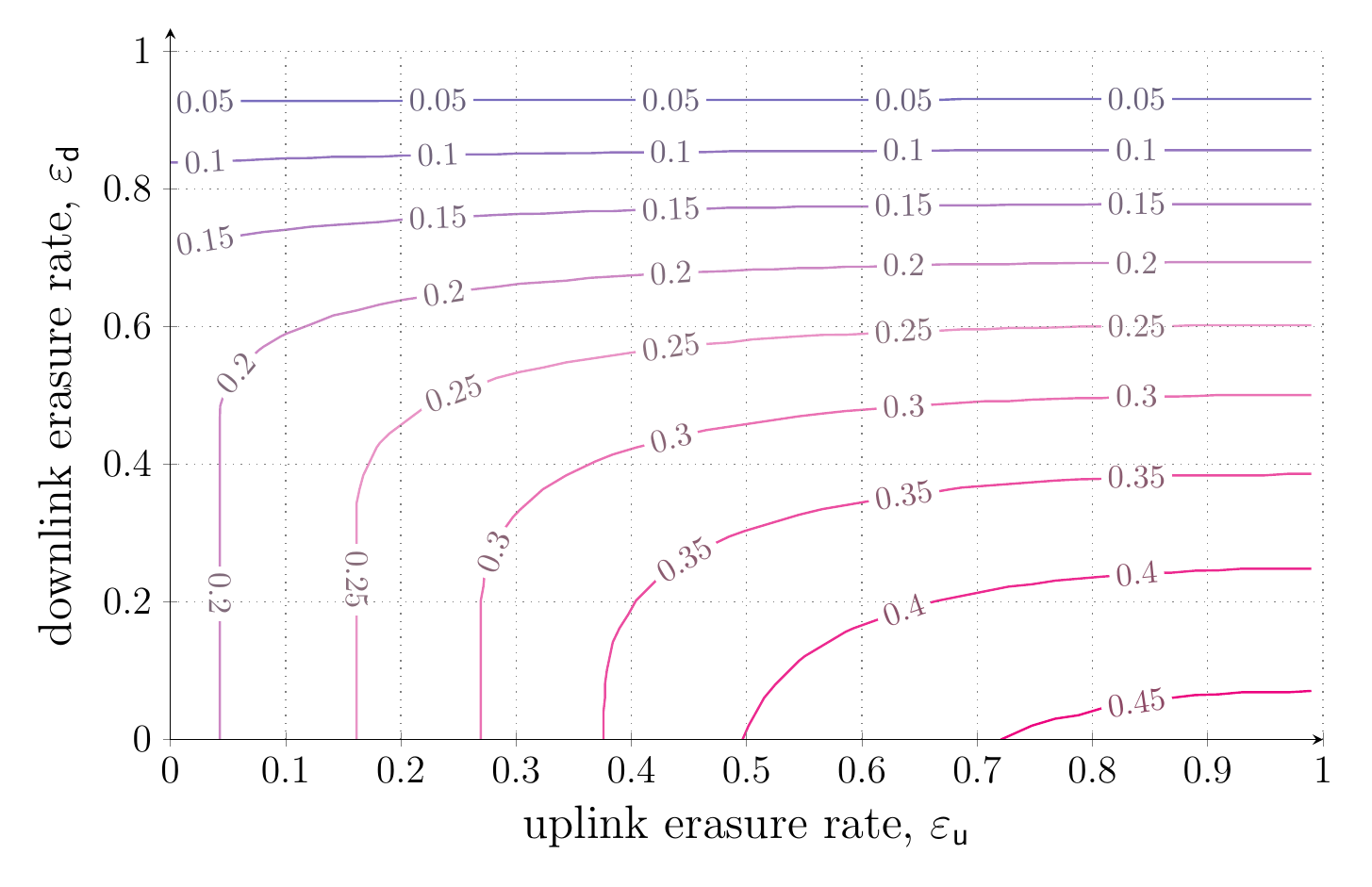}
\caption{Maximum achievable end-to-end throughput \maxTru\ for all configurations of \perasU\ and \perasD. $\nRx=2$, $\load=1/(1-\perasU)$.}
\label{fig:contour}
\end{figure}

We conclude our discussion of the two-relay setup considering the situation in which up- and down-link are characterised by distinct erasure rates. To this aim, Fig.~\ref{fig:contour} reports contour lines of \maxTru\ for all possible $(\perasU,\perasD)$ pairs, obtained via \eqref{eq:optTru}. In accordance with the trends observed thus far, the overall performance for $\load = 1/(1-\perasU)$ improves for higher values of \perasU\ \--- increasing the likelihood for at least one of the relays to decode \--- and for lower erasure rates on the downlink \--- favouring successful delivery to the sink once the forwarding probability has been optimised.

\begin{figure}
\centering
\includegraphics[width=.9\columnwidth]{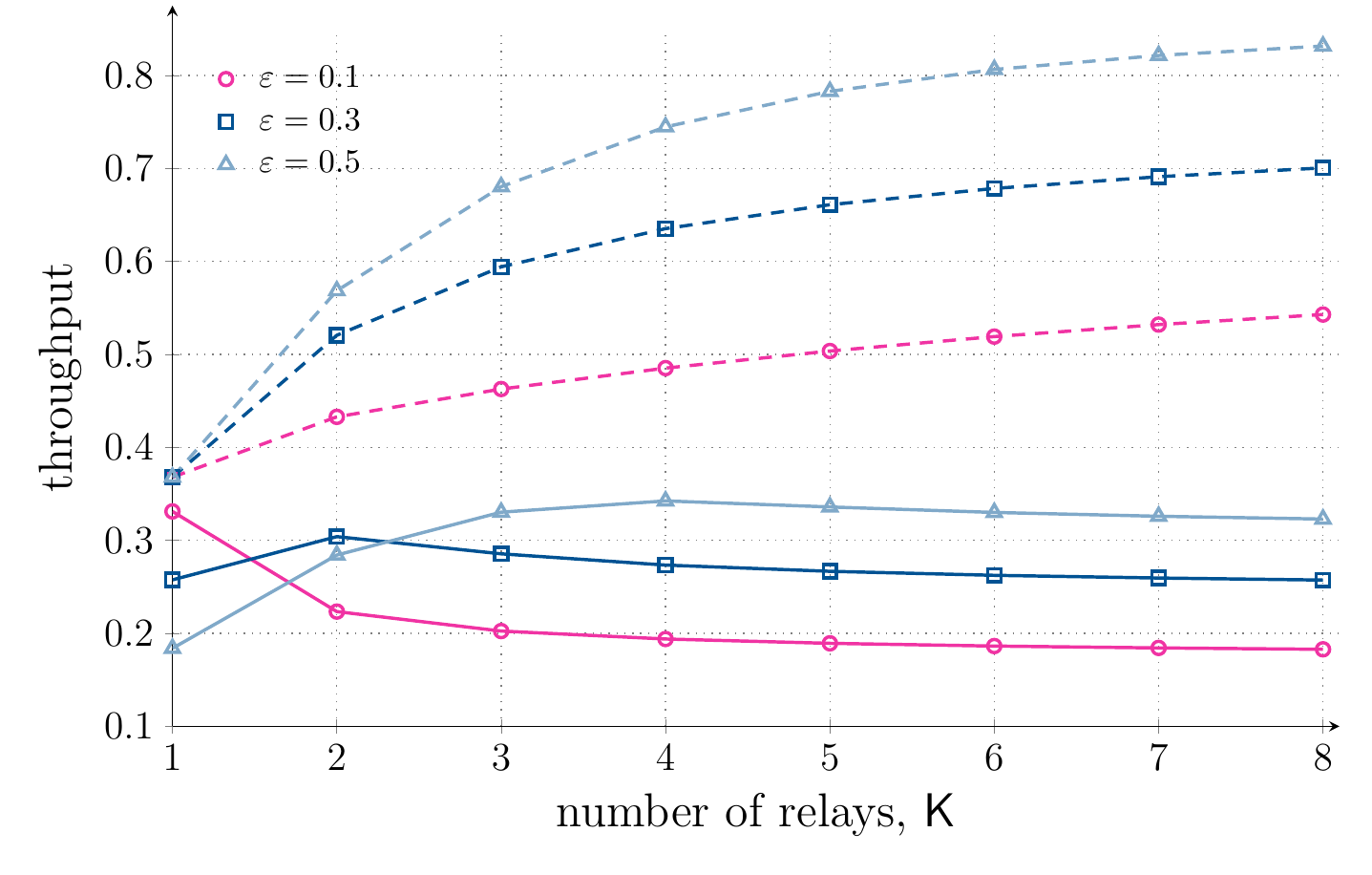}
\caption{Maximum achievable end-to-end throughput \maxTru\ (solid lines) and throughput upper bound \truB\ (dashed lines) vs number of receivers. Different markers report the behaviour of the system for different $\perasU=\perasD=\peras$. In all cases, $\load =1/(1-\peras)$.}
\label{fig:truKRxSymm}
\end{figure}

\subsection{The $\nRx >2$ case}

Fig.~\ref{fig:PeakTru2RxSymm} highlighted that the presence of multiple relays does not necessarily trigger end-to-end performance gains over the single-receiver configuration when the downlink is operated following a \ac{SA} policy. To gather a deeper understanding on this aspect, we report in Fig.~\ref{fig:truKRxSymm} the optimal throughput \maxTru\ (solid lines) and the upper bound \truB\ (dashed lines) against the number of relays \nRx, assuming $\perasU = \perasD = \peras$ and operating the uplink at load $\load = 1/(1-\peras)$. In the plot, curves labelled with the same marker refer to results obtained for the same erasure rate \peras. Albeit conceptually simple, the derivation of \maxTru\ for an arbitrary value of \nRx\ \--- equivalent to finding the maximum of a polynomial of order \nRx\ \--- does not lend itself to simple closed form expressions as the one reported in \eqref{eq:optTru} for the two-relay case. The reported trends have thus been obtained via numerical maximisation of \eqref{eq:tru} with respect to \pforw.

Under the ideal downlink conditions epitomised by the bound, an increase in \nRx\ is always favorable. Indeed, adding a receiver raises the probability for at least one packet to be collected over an uplink slot, with stronger improvements experienced for larger values of \peras\ in view of the more decorrelated reception pattern among the relays. When medium access contention and erasures in forwarding data units towards the sink come into play, however, the trend changes sharply, and the existence of an optimal number of receivers is apparent in Fig.~\ref{fig:truKRxSymm}. For low erasure rates, as already discussed in Fig.~\ref{fig:PeakTru2RxSymm}, a single-relay topology grasps better end-to-end throughput. Conversely, for larger values of \peras, the tradeoff between an higher likelihood of retrieving data over the uplink and the subsequently increased congestion in the donwlink benefits from receiver diversity (e.g. the best \maxTru\ is achieved for $\nRx=2$ if $\peras=0.3$ and for $\nRx=4$ if $\peras=0.5$).

From this standpoint, the presented framework offers compact and useful expressions to optimise system design, and is meant to stimulate further research on the potential of multi-receiver topologies for \ac{IoT} applications that rely on \ac{RA}.

\section{Conclusions}
\label{sec:conclusions}

In this paper, we analysed a two-tier system in which multiple receivers collect packets sent from users, and forward retrieved data units towards a sink. Both users-to-receivers and receivers-to-sink channels are operated following a \ac{SA} policy and are subject to packet erasures. In this setup, we derived exact expressions for the end-to-end throughput, and discussed how performance is influenced by channel impairments as well as by the cardinality of the relay set. We considered the possibility for receivers to probabilistically drop collected packets, and leveraged this to optimise the system behaviour. The analysis revealed interesting tradeoffs, clarifying how the number of receivers shall carefully be tuned based on channel and load parameters so as to maximise end-to-end throughput. The presented results call for further studies on how to reap the potential of such topologies, which can be of practical relevance for \ac{IoT} applications in beyond-5G systems. 
\appendices
\section{}
\label{app:tru}

In order to compute the end-to-end throughput, let us indicate for compactness $\pforw (1-\perasU) (1-\perasD) := \beta$. Recalling the definitions of \psU\ and \psD, \eqref{eq:truSum} can be written as
\begin{align}
\begin{split}
\tru &= \sum_{\nTx=0}^{\infty} \frac{\load^\nTx e^{-\load}}{\nTx!} \cdot \nRx \, \beta \, \nTx \perasU^{\nTx-1} \left( 1 - \beta \, \nTx \perasU^{\nTx-1}\right)^{\nRx-1}\\
&\stackrel{(a)}{=}\sum_{i=0}^{\nRx-1}(-1)^i \, \nRx {\nRx-1 \choose i} \frac{\beta^{i+1} \, e^{-\load}}{\perasU^{i+1}} \sum_{\nTx=0}^{\infty} \frac{\left(\load \, \perasU^{i+1}\right)^\nTx}{\nTx!} \cdot \nTx^{i+1}
\end{split}
\label{eq:truDerivation}
\end{align}
where $(a)$ follows by applying Newton's binomial expansion and after some simple yet tedious rearrangements. Let us now introduce the ancillary function
\begin{equation}
\ancF_{m}(x) := \sum_{\nTx=0}^{\infty} \frac{x^\nTx \,\nTx^m}{\nTx!}.
\end{equation}
By definition, $\ancF_0(x) = e^x$. Moreover, for $m\geq1$, we have
\begin{align}
\ancF_m(x) &= x \sum_{\nTx=0}^{\infty} \frac{x^{\nTx-1} \,\nTx^{m-1}}{(\nTx-1)!} \stackrel{(b)}{=} \,\,x \sum_{t=0}^{\infty} \frac{x^{t} \,(t+1)^{m-1}}{t!} \\
&\stackrel{(c)}{=} x \sum_{\ell=0}^{m-1} {m-1 \choose \ell}\sum_{t=0}^{\infty} \frac{x^{t} \,t^{\ell}}{t!}\\
&= x \sum_{\ell=0}^{m-1} {m-1 \choose \ell}\sum_{t=0}^{\infty} \ancF_\ell(x)
\end{align}
where $(b)$ applies the change of variable $t:=\nTx-1$ and $(c)$ results from applying once more Newton's binomial expansion to $(t+1)^{m-1}$, leading to the recursive definition of \eqref{eq:ancFunc}. Plugging this result into the innermost summation within \eqref{eq:truDerivation} leads to the sought expression of the end-to-end throughput reported in \eqref{eq:tru}. 

\section{}
\label{app:truBound}

By definition, the benchmark throughput \truB\ can be derived computing the probability that at least one of the relays decodes a packet over a slot. In turn, due to the independence of erasure events, the number of successful uplink receptions per slot conditioned on having $\NTx=\nTx$ follows a binomial distribution of parameters $(\nRx,\psU)$. Therefore, recalling the expression of \psU,
\begin{align}
\truB &= 1- \sum_{\nTx=0}^{\infty} \frac{\load^\nTx e^{-\load}}{\nTx!} \cdot \left[\, 1 - \nTx (1-\perasU) \perasU^{\nTx-1} \, \right]^{\nRx} \\
&\stackrel{(a)}{=} 1-\sum_{\ell=0}^{\nRx} (-1)^\ell {\nRx \choose \ell} \left(\frac{1-\perasU}{\perasU}\right)^{\ell} e^{-\load} \, \sum_{\nTx=0}^{\infty}\frac{\left( \load \, \peras^\ell\right)^{\nTx}}{\nTx!} \, \nTx^\ell
\end{align}
where $(a)$ resorts to Netwon's binomial expansion of $\left[\, 1 - \nTx (1-\perasU) \perasU^{\nTx-1} \, \right]^{\nRx}$. The expression in \eqref{eq:truBound} follows noting that the innermost summation is, by definition, $\ancF_\ell(\load\, \perasU^\ell)$.

\bibliographystyle{IEEEtran}
\bibliography{IEEEabrv,references,aloha}
\end{document}